\documentclass{article}

\usepackage{arxiv}

\usepackage[utf8]{inputenc} 
\usepackage[T1]{fontenc}    
\usepackage{hyphenat}
\usepackage{scrextend} 
\usepackage{hyperref}       
\usepackage{url}            
\usepackage{booktabs}       
\usepackage{longtable}
\usepackage{multirow}
\usepackage{array}
\usepackage{amsfonts}       
\usepackage{nicefrac}       
\usepackage{lipsum}
\usepackage{graphicx}
\usepackage{amsmath} 
\usepackage{amssymb}
\usepackage[dvipsnames,table,xcdraw]{xcolor}
\usepackage[multiple]{footmisc}
\usepackage{float}
\usepackage{soul}
\usepackage[noabbrev]{cleveref} 
\usepackage{caption} 
\usepackage{subcaption} 
\usepackage{textcomp}
\usepackage{enumitem}
\usepackage{comment} 

\usepackage{color}
\usepackage{authblk} 

\usepackage{listings}
\definecolor{codegreen}{rgb}{0,0.6,0}
\definecolor{codegray}{rgb}{0.5,0.5,0.5}
\definecolor{codepurple}{rgb}{0.58,0,0.82}
\definecolor{backcolour}{rgb}{0.95,0.95,0.95}

\lstdefinestyle{mystyle}{
    backgroundcolor=\color{backcolour},   
    commentstyle=\color{codegreen},
    keywordstyle=\color{magenta},
    numberstyle=\tiny\color{codegray},
    stringstyle=\color{codepurple},
    basicstyle=\ttfamily\footnotesize,
    breakatwhitespace=true,         
    breaklines=true,                 
    keepspaces=true,                 
}

\newcommand\inlinecode[1]{%
  \colorbox{backcolour}{\lstinline[basicstyle=\ttfamily\footnotesize]{#1}}%
}

\graphicspath{ {./images/} }

\newcommand{\nno}{\ensuremath{\mathrm{N_2O}}}
\newcommand{\no}{\ensuremath{\mathrm{NO_3}}}
\newcommand{\nh}{\ensuremath{\mathrm{NH_4}}}
\newcommand{\po}{\ensuremath{\mathrm{PO_4}}}
\newcommand{\coo}{\ensuremath{\mathrm{CO_2}}}

\definecolor{measurement}{HTML}{DAE8FC} 
\definecolor{controlled}{HTML}{FFFFC7}  
\definecolor{watchdog}{HTML}{B0E2AF}    
\definecolor{estimated}{HTML}{FFCE93} 

\usepackage[
    backend=biber,
    style=nature, 
    natbib
    ]{biblatex}

\title{Time Series Dataset for Modeling and Forecasting of \nno{} in Wastewater Treatment}

\author[1, 2, *]{ Laura Debel Hansen}
\author[1]{Anju Rani}
\author[2]{Mikkel Algren Stokholm-Bjerregaard}
\author[2]{Peter Alexander Stentoft}
\author[1]{Daniel Ortiz Arroyo}
\author[1]{Petar Durdevic}

\affil[1]{\footnotesize Department of Energy, Aalborg University, 6700 Esbjerg, Denmark}
\affil[2]{\footnotesize Krüger Veolia, Veolia Water Technologies, 2860 Søborg, Denmark}
\affil[*]{\footnotesize \texttt{laura.debel@veolia.com}  or  \texttt{ldh@energy.aau.dk}}

\begin{document}

\maketitle
\begin{abstract}
In this paper, we present two years of high-resolution nitrous oxide (\nno{}) measurements for time series modeling and forecasting in wastewater treatment plants (WWTP). 
The dataset comprises frequent, real-time measurements from a full-scale WWTP, with a sample interval of 2 minutes, making it ideal for developing models for real-time operation and control. 
This comprehensive bio-chemical dataset includes detailed influent and effluent parameters, operational conditions, and environmental factors. Unlike existing datasets, it addresses the unique challenges of modeling \nno{}, a potent greenhouse gas, providing a valuable resource for researchers to enhance predictive accuracy and control strategies in wastewater treatment processes. 
Additionally, this dataset significantly contributes to the fields of machine learning and deep learning time series forecasting by serving as a benchmark that mirrors the complexities of real-world processes, thus facilitating advancements in these domains. We provide a detailed description of the dataset along with a statistical analysis to highlight its characteristics, such as nonstationarity, nonnormality, seasonality, heteroscedasticity, structural breaks, asymmetric distributions, and intermittency, which are common in many real-world time series datasets and pose challenges for forecasting models.

\end{abstract}


\clearpage

\section*{Specifications}

\begin{table}[ht]
\renewcommand{\arraystretch}{1.5}
\centering
\label{tab:my-table}
\begin{tabular}{lp{11.5cm}}
\toprule
Subject                        & Environmental science and engineering                                                                                                       \\
\rowcolor[HTML]{F5F5F5} 
Specific subject area          &  Datadriven or mechanistic modelling or forecasting of \nno{} concentration in wastewater treatment plants utilizing the activated sludge process (ASP).                        \\
Type of data                   & Time series (csv).                                                                                                               \\
\rowcolor[HTML]{F5F5F5} 
Data acquisition &
The plant is controlled by an internal SCADA system, which receives setpoints from an online cloud-based platform developed by Krüger Veolia. 
Operational data was extracted from the online cloud platform at a sample interval of 2 min.
\\
Data format                    & 2-minute time-aggregated raw data                                                                                                      \\
\rowcolor[HTML]{F5F5F5} 
Description of data collection & 
Operational data was collected over 2 years from June 2022 to June 2024 using online chemical measurements of \nh{}, \no{}, \po{}, \nno{} and DO. Additionally, control signals generated in the SCADA system and cloud-based control platform was collected simultaneously alongside temperature, suspended solids, and flow measurements of both air and wastewater delivered to the process. 12 sensor measurements and 12 various other signals yields a total of 24 signals included in the dataset. \\
Data source location           & \begin{tabular}[c]{@{}l@{}}Institution: BIOFOS \\ City/region: Avedøre, Copenhagen \\ Country: Denmark\end{tabular} \\
\rowcolor[HTML]{F5F5F5} 
Data accessibility             & \begin{tabular}[c]{@{}l@{}}Repository name: Mendeley Data \\ DOI: 10.17632/xmbxhscgpr.1 \\ Direct URL to data: \small \url{https://data.mendeley.com/datasets/xmbxhscgpr/1} \end{tabular}          \\
Related Research            & Hansen, L. D., Stentoft, P. A., Ortiz-Arroyo, D. \& Durdevic, P. Exploring data quality and seasonal variations of N2O in wastewater treatment: a modeling perspective. Water Practice \& Technology, wpt2024045 (2024) \cite{hansen2024exploring} \\
\bottomrule
\end{tabular}
\end{table}

\section*{Value of the data}
\begin{itemize}
    \item Long-term high-resolution time series data from an operating full-scale wastewater treatment plant (WWTP) can help assess the factors and conditions that impact the \nno{} production in activated sludge processes. 
    \item Stakeholders in the water environments may take more informed decisions to optimally monitor, report, and control wastewater processes based on real-world operational data from an example case.
    \item Research studies in \nno{} model development, control development and mitigation strategies can benefit from access to real-world operational data.
    
\end{itemize}

\section{Objective}
Wastewater treatment plants (WWTPs) play a crucial role in modern society by treating polluted water from industrial and domestic sources before releasing the treated effluent into the environment. 
During the biological treatment process, WWTPs are known to produce significant amounts of nitrous oxide (\nno{})\citep{Law2012}. With a global warming potential 265 times that of \coo{} over a 100-year period \citep{IPCC2013}, \nno{} is a potent greenhouse gas (GHG) and a major contributor to the climate footprint of WWTPs. Consequently, \nno{} has gained attention in the wastewater industry and political discussions, making mitigation a priority for WWTPs.
 
Understanding the \nno{} production mechanisms and modelling of it in full-scale WWTPs is crucial for mitigation of \nno{} on a global level. 
To support the development of models and mitigation strategies, this time series dataset has been collected, encompassing various operational conditions during 2 years of operation in a full-scale WWTP.

The dataset serves as a benchmark for evaluating the performance of different modelling and mitigation techniques. This dataset aims to foster collaboration and knowledge sharing among industry experts, researchers, and stakeholders involved in the water sector and beyond. 
Additionally, the availability of this dataset can encourage the adoption of standardized testing protocols and best practices for modelling wastewater treatment processes in general. 
By utilizing the dataset, researchers can explore the effect of applying a wide range of data-driven and mechanistic methodologies. The dataset promotes collaboration, standardization, and innovation in the modelling, control, and mitigation of \nno{}, facilitating the adoption of advanced techniques and best practices across industries.

\section{Materials and Methods}
\subsection{Plant description}
\begin{figure}[ht]
    \centering
    \includegraphics[width=0.9\textwidth]{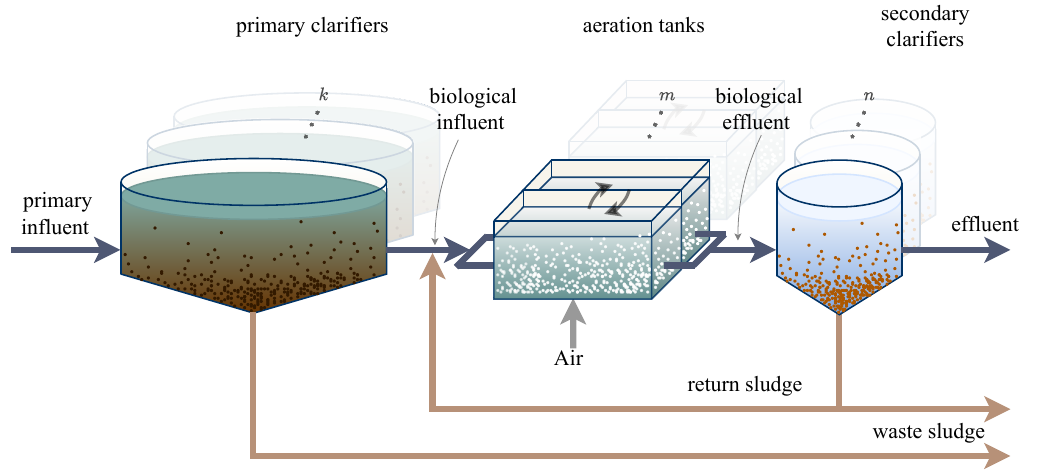}
    \caption{Layout of a WWTP with ASP. The parameters $k$, $m$, and $n$ vary based on plant-specific dimensioning. Adapted from \citep{hansen2024exploring}.}
    \label{fig:wwtp}
\end{figure}

\begin{figure}[ht]
    \centering
    \includegraphics[width=0.9\textwidth]{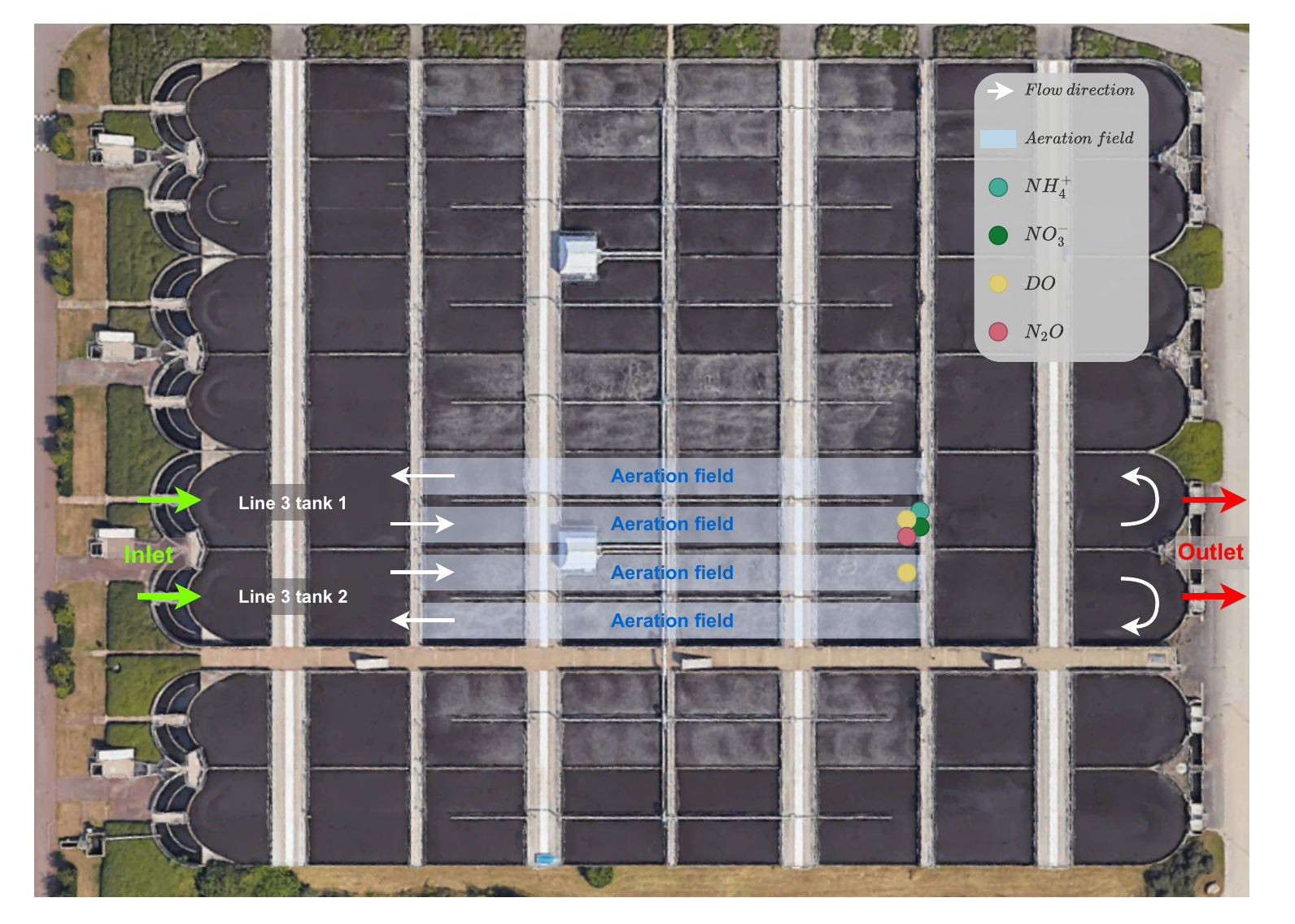}
    \caption{Aerial photograph of the 8 aeration tanks at the case plant. The sensor placement, aeration field and flow directions are indicated in tank 1 and tank 2 for line 3.}
    \label{fig:sensors}
\end{figure}

The data is collected from Avedøre WWTP, located in the urban area of Copenhagen, Denmark. Biological treatment is applied in four parallel lines with two carousel reactors in each line. The process of interest is the activated sludge process (ASP), which occurs in the aeration tanks as shown in \cref{fig:wwtp}. Here, microbial treatment in the form of nitrification and denitrification treatment takes place by controlling the presence of air. 
Alternating ASP is achieved with bubble diffusers mounted at the bottom of each reactor and is controlled based on the concentration of ammonium (\nh{}), nitrate (\no{}), and dissolved oxygen ($O_2$) concentration. 
After the treatment, the activated sludge is separated from the mixed liquor and recycled into the system to continue the treatment process.
The data is collected from line 3 with most sensors located in one carousel reactor.
Other operational variables that are monitored suspended solids (SS), temperature (T), phosphate ($PO_4$), valve position in aeration pipe, airflow from the joint blower station, inflow of wastewater, and control signals.
Additionally, the airflow to each tank is estimated based on the valve position and airflow from the joint blower station. 
\Cref{fig:sensors} illustrates the location of chemical sensors.

The case plant treats wastewater for 350.000 population equivalent (PE). Operational data related to the biological process in the aeration tank is available from June 2022 to June 2024.

\begin{table}[ht]
\centering
\caption{Categorized sensor data and control signals available in the dataset. The column \textit{Add. location} indicates if the signal is also available at another location. 
The row color indicates: 
\colorbox{measurement}{measurement (blue)}, 
\colorbox{estimated}{estimated values (orange)},
\colorbox{controlled}{controlled states (yellow)}, and 
\colorbox{watchdog}{watchdog (green)}. }
\label{tab:data_types}
\resizebox{\textwidth}{!}{%
\begin{tabular}{llllp{6cm}}
\toprule
Name &
  Add. location &
  Abbrv. &
  Unit &
  Comment \\ \bottomrule
\rowcolor{watchdog} INLET.STATE.SWM INLET FLOW &
   &
  SWM &
  - &
  Storm Water mode state. An integer signal describing if the plant runs a special wet-weather control due to high inflow usually caused by precipitation. \\
\rowcolor{measurement} INLET.Q &
   &
  $Q_{in}$ &
  m$^3$/h &
  Total flow of wastewater \\
\rowcolor{measurement}BIOLOGY.BLOWERSTATION 1.Q.AIRFLOW &
   &
  $Q_{air,total}$ &
  Nm$^3$/h &
  Combined airflow from blower station \\
\rowcolor{measurement} BIOLOGY.LINE 3 TANK 1.N2O &
   &
  \nno &
  mg-N/L &
  Nitrous Oxide \\
\rowcolor{measurement} BIOLOGY.LINE 3 TANK 1.NH4 &
   &
  \nh &
  mg-N/L &
  Ammonium \\
\rowcolor{measurement} BIOLOGY.LINE 3 TANK 1.NO3 &
   &
  \no &
  mg-N/L &
  Nitrate \\
\rowcolor{measurement} BIOLOGY.LINE 3 TANK 1.O2 &
  TANK 2 &
  $\text{O}_2$ &
  mg/L &
  Dissolved Oxygen \\
\rowcolor{measurement} BIOLOGY.LINE 3 TANK 1.SS &
  TANK 2 &
  SS &
  g/L &
  Suspended solids \\
\rowcolor{measurement} BIOLOGY.LINE 3 TANK 1.PO4 &
   &
  $PO_4$ &
  mg-P/L &
  Phosphate \\
\rowcolor{measurement} BIOLOGY.LINE 3 TANK 1.TEMPERATURE &
  TANK 2 &
  T &
  $^\circ$C &
  Temperature \\
\rowcolor{estimated} BIOLOGY.LINE 3 TANK 1.Q.AIRFLOW &
  TANK 2 &
  $Q_{air}$ &
  Nm$^3$/h &
  Estimated Airflow over valve \\
\rowcolor{controlled} BIOLOGY.LINE 3 TANK 1 VALVE 1.PCT &
   TANK 2 &
  $u_{v}$ &
  \% &
  Valve position \\
\rowcolor{controlled} BIOLOGY.LINE 3.PHASECODE.SETPOINT &
   &
  $\phi$ &
  - &
  Phase code. A line-specific 4-digit code indicating the inlet and outlet tank, along with the aeration phase in each tank. \\
\rowcolor{controlled} BIOLOGY.LINE 3 TANK 1.PROCESSPHASE &
   TANK 2 &
  $\phi_{1}$,$\phi_{2}$ &
  - &
  Aeration phase. An integer signal describing conditions in the given tank. \\
\rowcolor{controlled} BIOLOGY.LINE 3.PROCESSPHASE.INLET TANK &
   &
  $\phi_{in}$ &
  - &
  Inlet tank. An integer signal describing if wastewater is flowing to tank 1 or 2 \\
\rowcolor{controlled} BIOLOGY.LINE 3.PROCESSPHASE.OUTLET TANK &
   &
  $\phi_{out}$ &
  - &
  Outlet tank. An integer signal describing if wastewater is exiting from tank 1 or 2 \\ 
  \bottomrule
\end{tabular}}
\end{table}

\subsection{Data Collection}
Data was collected using the online cloud platform, Hubgrade Performance Plant\footnote{\url{ https://www.veoliawatertechnologies.com/en/hubgrade-performance-plant-module-digital-solution}} (HPP), connected to the plant's SCADA system. As this is operational data, no additional procedures were implemented to acquire it. The sensors were maintained according to standard operational practices, without extra measures to enhance data quality beyond what is typical for a Danish WWTP. This ensures a comprehensive dataset representing real world conditions with challenges such as missing data due to maintenance and replacement of equipment. 
The sample period ranges between 2-5 minutes, with all data subsequently up-sampled to preserve the signal dynamics. The information is categorized into several types: measurements of ongoing processes, control signals generated by the SCADA system, alarm signals (or watchdogs) that trigger operational mode shifts upon exceeding predefined thresholds, and process modes, represented by Boolean signals or integer codes indicating the current operational state. 
There are 16 different types of signals which are all described in table \ref{tab:data_types}. Table \ref{tab:data_types} furthermore provides an overview of the system measurements, controlled variables, watchdog and estimated variables.

\subsubsection{N2O sensor}
Dissolved \nno{} concentrations were measured using a Clark-type electro-chemical \nno{} sensor from Unisense Environment, Denmark\citep{UnisenseN2O}. The sensor features a replaceable sensor head with a manufacturer-guaranteed lifespan of 4 months, though a lifespan of over 6 months is expected. 
Prior to deployment, the sensor is calibrated using a two-point calibration at the same temperature as the wastewater. This ensures measurements are accurate within an uncertainty of $\pm 5\%$ within $\pm 3^\circ C$  of the calibration temperature. To account for seasonal temperature variations, it is recommended to perform a two-point calibration every two months. 
However, it should be noted that no metadata of performed calibrations are available, which is an issue addressed is \cite{hansen2024exploring}.
The sensor measures, in its default configuration,\nno{} concentrations within the range of 0-1.5 mg \nno{}-N/L with a resolution of 0.005 mg \nno{}-N/L. However, the sensor can be recalibrated to suit the anticipated \nno{} concentration range at a specific plant, although this adjustment may reduce resolution, resulting in larger measurement increments. 
The sensor's response time is 65 seconds for temperatures between 10-30$^\circ C$, with slower response times at lower temperatures. Information about the sensor was obtained from the manufacturer's manual and personal communication.

\subsubsection{Airflow estimation}
The plant has a shared compressor station that delivers an airflow to a shared manifold. The valve opening to each tank then determines the flow to each specific tank. An approximation of the airflow to each tank is given in \cref{eq:airflow_correaction_factor,eq:airflow_estimate}, using the total airflow delivered by the compressor and valve opening to the specific tank.

The airflow correction factor, $C_{v,i}$ based on the valve position, $U_{v,i}$:
\begin{equation}
\label{eq:airflow_correaction_factor}
    C_{i}(U_{v,i}) =  
\begin{cases}
    \frac{60}{33} U_{v,i} ,   & \text{if } 0 \leq U_{v,i} \leq 33\\
    \frac{30}{33} U_{v,i} ,   & \text{if } 33 < U_{v,i} \leq 66\\
    \frac{10}{34} U_{v,i} ,   & \text{if } 66 < U_{v,i} \leq 100\\
\end{cases}
\end{equation}

Here, $C_{v,i}$, is an airflow correction factor based on the valve position, $U_{v,i}$.  As soon as a valve is open, it takes a major part of the combined airflow. However, the airflow over the valve does not increase linearly with the valve position, but rather in a sigmoid-shaped way as given by \cref{eq:airflow_correaction_factor}. Using this relationship, the most open valve will have the highest estimated airflow. 
The estimated airflow to tank $i$ is given in \cref{eq:airflow_estimate}. 

\begin{equation}
\label{eq:airflow_estimate}
    \widehat{Q}_{air,i} = \frac{C_{i}(U_{v,i})}{\sum\limits_{j=1}^{n} C_{j}(U_{v,j})}  Q_{air,total}
\end{equation}

in which $\widehat{Q}_{air,i}$ is the estimated airflow to tank $i$. $C_{i}(U_{v,i})$ is the flow correction factor as a function of the valve position $U_{v,i}$. The total airflow delivered by the combined blower station is denoted $Q_{air,total}$.

\subsubsection{Variable naming}
All variables in the dataset are named according to their location and the type of observation; sensor measurement, setpoints, watchdogs and process modes. The variable name is composed by  specifications separated by a point (.) as described here: 
\begin{lstlisting}[numbers=none]
name = {UnitOperation}.{SubUnitOperation}.{Type}.{SubType} 
\end{lstlisting}

The different specification elements in the name are listed below:
\begin{lstlisting}[numbers=none]
UnitOperation: 
    The main location specification which indicates if the variable is collected at the inlet to the WWTP or in the biological process. Can be "Inlet" or "Biology".
SubUnitOperation: 
    The second location specification (if any) which specifies the location within the unit. Can specify the line number, tank number, "blower station", "state" or "process phase" 
Type: 
    The main type specification, such as "N2O", "NH4", "Q", etc.
SubType: 
    Specifies the sub-type such as "setpoint". 
\end{lstlisting}
This variable naming convention is also used in \cref{tab:data_types}.

\subsubsection{Phase codes and control inputs}
All biological processes (nitrification, denitrification) are controlled using integer code to describe the conditions governing each biological tank. A four-digit code is used in the online control system to describe the process for each of the biological lines shown in \cref{fig:sensors}. These codes are referred to as biological phase codes. The meaning of each digit in the phase codes is described in \cref{tab:phasecodes}.

\begin{table}[h!]
\centering
\caption{Biological phase codes definition.}
\label{tab:phasecodes}
\begin{tabular}{lll}
\toprule
$\phi$ & Description & Value\\
\midrule
$\phi_{in}$  & The tank that wastewater flows into &Takes a value of 1 or 2 \\
$\phi_{out}$  & The tank that effluent flows from &Takes a value of 1 or 2 \\
$\phi_{1}$  & Conditions in tank 1 &Takes a value of 0, 1, or 2   \\
$\phi_{2}$  & Conditions in tank 2 &Takes a value of 0, 1, or 2   \\
\bottomrule
\end{tabular}
\end{table}

As noted in \cref{tab:phasecodes}, digit $\phi_1$ and $\phi_2$ can take values between 0-2. They define the conditions in tank 1 and 2, respectively. The three possible options are:

\begin{labeling}{2:} 
  \setlength\itemsep{-0.1cm}
    \item[0:] Anaerobic conditions without aeration and mixing.
    \item[1:] Anoxic conditions (denitrification) without aeration and with mixing. 
    \item[2:] Aerobic conditions (nitrification) with aeration and mixing. 
\end{labeling}

\textbf{Example} \\
A phasecode $\phi = 1112$ means: inflow to tank 1, outflow from tank 1, denitrification in tank 1 and nitrification in tank 2. \\
A phasecode $\phi = 1222$ means: inflow to tank 1, outflow from tank 2, nitrification in tank 1 and nitrification in tank 2. \\
A phasecode $\phi = 2222$ means: inflow to tank 2, outflow from tank 2, nitrification in tank 1 and nitrification in tank 2. 

\subsubsection{Quality Assessment}

Each signal in the dataset is accompanied by a corresponding quality signal that evaluates its quality at every time step. The quality is indicated by an integer, following this rule:
\begin{lstlisting}[numbers=none]
0 = Good quality 
1 = Bad quality 
\end{lstlisting}
A straightforward quality assessment is performed for each signal collected in the HPP. A measurement can be classified as having poor quality due to two reasons: (1) the sensor itself indicates poor quality, typically due to maintenance issues as specified by the manufacturer, or (2) the value remains constant over a fixed period, at which point the HPP flags the signal as corrupted.

A watchdog, setpoint, or control state is considered to have poor quality if it is not utilized in the plant's online control system. Similarly, an estimate's quality is considered poor if any of the signals used in its calculation were given poor quality.

The quality assessment is included as a separate variable, resulting in two column names for each measurement. An example is given here for the \nno{} measurements: \inlinecode{BIOLOGY.LINE 3 TANK 1.N2O value} and \inlinecode{BIOLOGY.LINE 3 TANK 1.N2O quality}. 
It is noteworthy that the \inlinecode{value} variable remains unaltered by the \inlinecode{quality} variable; therefore, users of the dataset who are interested in conducting their own quality assessment can do so, as the values represent raw, unfiltered data.

\subsection{Data Visualization and Summary Statistics}

\begin{figure}[ht]
    \centering
    \includegraphics[width=\textwidth]{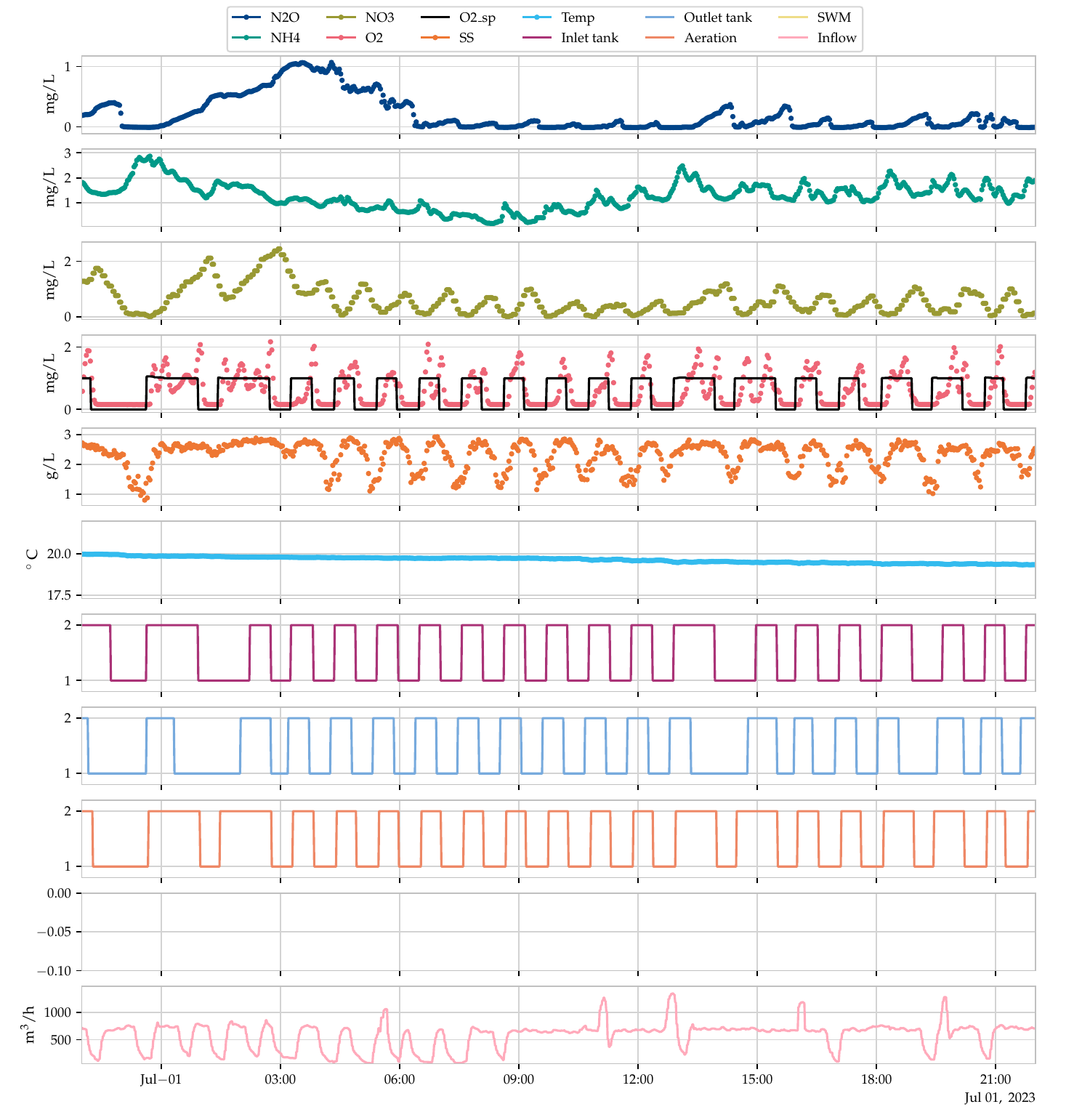}
    \caption{Example of variations in \nno{} concentration and other signals over an arbitrary period of 24 hours.}
    \label{fig:timeseries_example_1day}
\end{figure}

\begin{figure}[ht]
    \centering
    \includegraphics[width=\textwidth]{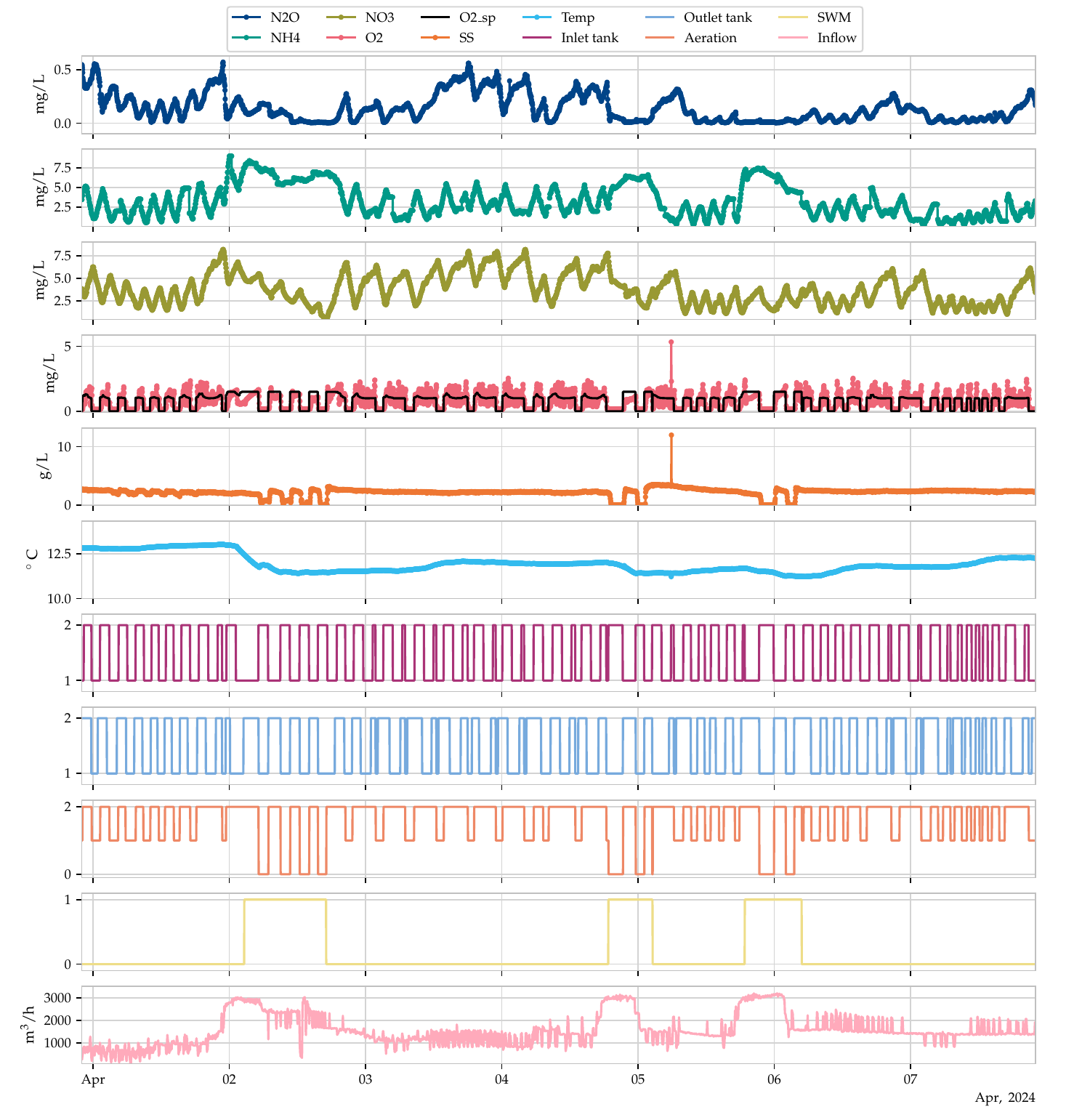}
    \caption{Example of variations in \nno{} concentration and other signals over an arbitrary period of 7 days.}
    \label{fig:timeseries_example_7days}
\end{figure}

\Cref{fig:timeseries_example_1day,fig:timeseries_example_7days} illustrate the hourly and daily variations in \nno{} concentation and other measurements and signals. These examples, taken from July 2023 (1 day) and April 2024 (7 days), respectively, demonstrate the dynamic changes in the process over different time frames.
\Cref{fig:timeseries_example_1day} highlights the impact of alternating control of the activated sludge process (ASP), showing the dissolved oxygen setpoint (O2\_sp) being regulated with varying phase durations. This variation is also evident in \cref{fig:timeseries_example_7days}, where the phase durations of the aeration, inlet tank, and outlet tank fluctuate over time.
Additionally, \cref{fig:timeseries_example_7days} provides several instances of storm water mode (SWM) activation due to heavy precipitation. In this special state, the system prioritizes preventing sludge from being discharged from the tanks, which may compromise nutrient removal.

The summary statistics are presented in \cref{tab:stats} and provide a comprehensive view of the data characteristics for each sensor measurement included in the dataset. For the control inputs and phase descriptions, i the statistics are a representation of the state of the system. 
Below is the explanation of each column in the table:

\begin{itemize}
    \item \textbf{Variable}: The name of the data column in the dataset.
    \item \textbf{Mean}: The average value of each variable.
    \item \textbf{Std}: The standard deviation of each variable.
    \item \textbf{Min}: The minimum value of each variable.
    \item \textbf{Q1}: The first quartile (25th percentile) of each variable.
    \item \textbf{Q2}: Median (50th percentile) of each variable.
    \item \textbf{Q3}: The third quartile (75th percentile) of each variable.
    \item \textbf{Max}: The maximum value of each variable.
\end{itemize}

\begin{table*}[ht]
\centering
\caption{Summary statistics of the raw data. The rows are color-coded to indicate the signal type: 
\colorbox{measurement}{measurement (blue)}, 
\colorbox{estimated}{estimated values (orange)},
\colorbox{controlled}{controlled states (yellow)}, and 
\colorbox{watchdog}{watchdog (green)}.}
\label{tab:stats}
\resizebox{\linewidth}{!}{
\begin{tabular}{lrrrrrrr}
\toprule
 Variable                                     & mean   & std    & min     & Q1     & Q2     & Q3     & max     \\ \hline
\rowcolor{measurement} 
INLET.Q value                                 & 3080.9 & 2329.9 & -5572.5 & 2095.5 & 2303.8 & 4329.9 & 17192.6 \\
\rowcolor{measurement}  
BIOLOGY.BLOWERSTATION 1.Q.AIRFLOW value & 17099.4 & 7333.1 & 0.0 & 11552.6 & 16754.5 & 21765.0 & 43356.6 \\
\rowcolor{measurement}  
BIOLOGY.LINE 3 TANK 1.N2O value               & 0.1    & 0.2    & -0.0    & -0.0   & 0.0    & 0.1    & 12.0    \\
\rowcolor{measurement}  
BIOLOGY.LINE 3 TANK 1.NH4 value               & 2.3    & 2.2    & -0.0    & 1.0    & 1.9    & 3.0    & 20.0    \\
\rowcolor{measurement}  
BIOLOGY.LINE 3 TANK 1.NO3 value               & 2.8    & 2.8    & -0.3    & 1.2    & 2.2    & 3.7    & 52.7    \\
\rowcolor{measurement}  
BIOLOGY.LINE 3 TANK 1.O2 value                & 0.6    & 0.6    & -0.0    & 0.2    & 0.3    & 1.0    & 10.9    \\
\rowcolor{measurement}  
BIOLOGY.LINE 3 TANK 1.SS value                & 2.6    & 0.9    & 0.0     & 2.1    & 2.5    & 3.0    & 35.1    \\
\rowcolor{measurement}  
BIOLOGY.LINE 3 TANK 1.TEMPERATURE value       & 15.7   & 3.4    & 6.5     & 12.6   & 15.7   & 18.9   & 27.1    \\
\rowcolor{measurement}  
BIOLOGY.LINE 3 TANK 1.PO4 value               & 1.1    & 1.0    & 0.0     & 0.5    & 0.9    & 1.4    & 16.6    \\
\rowcolor{measurement}  
BIOLOGY.LINE 3 TANK 2.O2 value                & 0.7    & 0.8    & -0.0    & 0.2    & 0.4    & 1.1    & 10.9    \\
\rowcolor{measurement}  
BIOLOGY.LINE 3 TANK 2.SS value                & 2.6    & 0.9    & 0.0     & 2.0    & 2.7    & 3.2    & 50.1    \\
\rowcolor{measurement}  
BIOLOGY.LINE 3 TANK 2.TEMPERATURE value       & 15.6   & 3.4    & 6.3     & 12.6   & 15.5   & 18.8   & 27.1    \\
\rowcolor{estimated}
BIOLOGY.LINE 3 TANK 1.Q.AIRFLOW value         & 2185.0 & 2131.2 & 0.0     & 0.0    & 2160.6 & 4008.8 & 12973.0 \\
\rowcolor{estimated}
BIOLOGY.LINE 3 TANK 2.Q.AIRFLOW value         & 2156.1 & 2091.7 & 0.0     & 0.0    & 2128.8 & 3903.5 & 12703.3 \\
\rowcolor{controlled} 
BIOLOGY.LINE 3 TANK 1 VALVE 1.PCT value       & 44.5   & 42.1   & 0.0     & 0.0    & 40.8   & 89.6   & 100.0   \\
\rowcolor{controlled} 
BIOLOGY.LINE 3 TANK 1.O2.SETPOINT value       & 0.6    & 0.6    & 0.0     & 0.0    & 1.0    & 1.0    & 2.5     \\
\rowcolor{controlled} 
BIOLOGY.LINE 3 TANK 1.PROCESSPHASE value      & 1.5    & 0.6    & 0.0     & 1.0    & 2.0    & 2.0    & 2.0     \\
\rowcolor{controlled} 
BIOLOGY.LINE 3 TANK 2 VALVE 1.PCT value       & 43.3   & 41.0   & 0.0     & 0.0    & 34.1   & 83.3   & 100.0   \\
\rowcolor{controlled} 
BIOLOGY.LINE 3 TANK 2.O2.SETPOINT value       & 0.7    & 0.6    & 0.0     & 0.0    & 1.0    & 1.1    & 2.5     \\
\rowcolor{controlled} 
BIOLOGY.LINE 3 TANK 2.PROCESSPHASE value      & 1.5    & 0.6    & 0.0     & 1.0    & 2.0    & 2.0    & 2.0     \\
\rowcolor{controlled} 
BIOLOGY.LINE 3.PHASECODE.SETPOINT value & NA & NA  & 0.0 & 1112.0  & 1222.0  & 2221.0  & 2222.0  \\
\rowcolor{controlled} 
BIOLOGY.LINE 3.PROCESSPHASE.INLET TANK value  & 1.5    & 0.5    & 0.0     & 1.0    & 1.0    & 2.0    & 2.0     \\
\rowcolor{controlled} 
BIOLOGY.LINE 3.PROCESSPHASE.OUTLET TANK value & 1.5    & 0.5    & 0.0     & 1.0    & 2.0    & 2.0    & 2.0     \\
\rowcolor{watchdog} 
INLET.STATE.SWM INLET FLOW value              & 0.1    & 0.3    & 0.0     & 0.0    & 0.0    & 0.0    & 1.0     \\ \bottomrule
\end{tabular}}
\end{table*}

\clearpage
\section*{Ethics Statement}
This research does not involve experiments, observations, or data collection related to human or animal subjects.

\section*{Declaration of competing interest}
The authors declare that they have no known competing financial interests or personal relationships that could have appeared to influence the work reported in this paper.

\section*{Data Availability}
\url{https://data.mendeley.com/datasets/xmbxhscgpr/1}

\section*{Acknowledgement}
We extend our sincere gratitude to Avedøre WWTP for their cooperation and for granting us permission to use and publish their data. Their support has been invaluable to this research.
This research was supported by Aalborg University and Krüger Veolia under the Danish Innovation Foundation grant number 1044-00031B.

\printbibliography

@techreport{IPCC2013,
    title = {{Climate Change 2013: The Physical Science Basis. Contribution of Working Group I to the Fifth Assessment Report of the Intergovernmental Panel on Climate Change}},
    year = {2013},
    author = {{IPCC}},
    pages = {1535},
    institution = {Cambridge University Press},
    address = {Cambridge, United Kingdom and New York, USA}
}

@article{hansen2024exploring,
    title = {{Exploring data quality and seasonal variations of N2O in wastewater treatment: a modeling perspective}},
    year = {2024},
    journal = {Water Practice {\textbackslash}{\&} Technology},
    author = {Hansen, Laura Debel and Stentoft, Peter Alexander and Ortiz-Arroyo, Daniel and Durdevic, Petar},
    pages = {wpt2024045},
    publisher = {IWA Publishing}
}

@misc{UnisenseN2O,
    title = {{N2O WASTEWATER SYSTEM}},
    year = {2022},
    author = {{Unisense Environment A/S}},
    url = {https://unisense-environment.com/wp-content/uploads/2022/11/2022.11-N2O-WW-System_english-2.pdf},
    institution = {Unisense Environment A/S}
}

@article{Law2012,
    title = {{Nitrous oxide emissions from wastewater treatment processes}},
    year = {2012},
    journal = {Philosophical Transactions of the Royal Society B: Biological Sciences},
    author = {Law, Yingyu and Ye, Liu and Pan, Yuting and Yuan, Zhiguo},
    number = {1593},
    pages = {1265--1277},
    volume = {367},
    doi = {10.1098/rstb.2011.0317},
    issn = {14712970}
}



\end{document}